\documentclass[12pt,preprint]{aastex}
\begin{document}

\title{Hubble Space Telescope STIS Spectroscopy of
Long Period Dwarf Novae in Quiescence}

\author{Edward M. Sion}
\affil{Department of Astronomy and Astrophysics, 800 Lancaster Ave.,
Villanova University, Villanova, PA 19085}
\email{\it emsion@ast.vill.edu}

\author{Boris T. G\"ansicke}
\affil{Department of Physics, University of Warwick, Coventry CV4 9BU, UK}
\email{\it Boris.Gaensicke@warwick.ac.uk}

\author{Knox S.Long}
\affil{Space Telescope Science Institute, 3700 San Martin Ave. 
Baltimore, MD 21218}
\email{\it long@stsci.edu}

\author{Paula Szkody}
\affil{Department of Astronomy, University of Washington, Seattle, WA 98195}
\email{\it szkody@astro.washington.edu}

\author{Christian Knigge}
\affil{Dept. of Physics and Astronomy, University of Southampton,
Highfield, Southamption, SO17 1BJ, UK}
\email{\it christian@astro.soton.ac.uk}

\author{Ivan Hubeny}
\affil{Steward Observatory and Department of Astronomy, University of Arizona, Tucson, AZ85721}
\email{\it hubeny@as.arizona.edu}

\author{Domitilla deMartino}
\affil{Osservatorio di Capodimonte, Via Moiariello 16, Naples I-80131, Italy}
\email{\it demartin@na.astro.it}

\author{Patrick Godon\altaffilmark{1}}
\affil{Department of Astronomy and Astrophysics
Villanova University, Villanova, PA 19085}
\email{patrick.godon@villanova.edu}

\altaffiltext{2}
{Visiting at the Space Telescope Science Institute,
Baltimore, MD 21218, godon@stsci.edu}

\begin{abstract}

We present the results of a synthetic spectral analysis of {\it{HST}} STIS
spectra of five long period dwarf novae obtained during their quiescence
to determine the properties of their white dwarfs which are little known
for systems above the CV period gap. The five systems, TU Men, BD Pav, SS
Aur, TT Crt, and V442 Cen were observed as part of an {\it{HST}} Snapshot
project. The spectra are described and fitted with combinations of white
dwarf photospheres and accretion disks. We provide evidence that the white
dwarfs in all five systems are at least partially exposed. We discuss the
evolutionary implications of our model fitting results and compare these
dwarf novae to previously analyzed FUV spectra of other dwarf novae above
the period gap. The dispersion in CV WD temperatures above the period gap
is substantially greater than one finds below the period gap where there
is a surprisingly narrow dispersion in temperatures around 15,000K.  
There appears to be a larger spread of surface temperatures in dwarf novae
above the period than is seen below the gap.

\end{abstract}


\section{Introduction}

Dwarf novae (DNe) are close interacting binaries in which a Roche-lobe
filling main sequence-like dwarf transfers matter with angular momentum
through a disk onto a white dwarf (WD). The rapid disk accretion during
outburst, due to a thermal instability that causes cyclic changes of the
accretion rate, releases gravitational potential energy identified as the
DN outburst.  The high accretion rate ($\sim 10^{-8}$ to $10^{-9}$
M$_{\odot}$/yr) outburst phase (which lasts a few days to weeks) is
preceded and followed by a low accretion rate ($\sim 10^{-11}
M_{\odot}$/yr)  quiescence stage. This DN behavior is punctuated every few
thousand years or more by episodes of explosively unstable thermonuclear
burning, the classical nova explosion (Warner 1995 and references therein).
Perhaps the least understood topic in CV/DN research (along with what drives 
the wind outflow in outburst) is the state and structure of the boundary layer 
and accretion disk during
quiescence and the physics of how long term accretion of mass, angular
momentum and energy affects the WD.  Our studies with archival {\it{IUE}},
and {\it{HST}} STIS have found that $\sim$50\% of the DNe in quiescence
are dominated (i.e., $>$ 60\% of UV flux) by the accretion disk;
$\sim$25\% are dominated by the WD and $\sim$25\% have nearly equal
contribution of WD and accretion disk (40-60\% each) \citep{urb06}

A number of studies \citep{sio91,ara05,urb06, tow03} have shown that CV
WDs above the gap are typically on-average $\sim 10,000$K hotter than CV
WDs below the period gap (almost certainly a consequence of higher
time-averaged accretion rates of systems above the gap but possibly with
system total age also being a factor).

It is also true that far fewer systems with reliably known WD properties
are known above the period gap compared with below the gap, thus impeding
detailed comparisons between the two groups. For example, among CVs below
the gap, there are now roughly 20 systems with reliable WD temperatures
compared with only 8 such systems above the gap. The primary reason for
this disparity is that in long period CVs, mass transfer rates are higher
and their disks brighter. Hence, it is difficult to disentangle the white
dwarf flux contribution from that of the accretion disk or other second
component of FUV flux. The key to disentangling the FUV components and
identifying the dominant component is the system distance.

As part of our effort to increase the sample of CV degenerates with known
properties above the gap, we have analyzed the {\it{HST}} STIS spectra of
five long period dwarf novae, BD Pav, TU Men, TT Crt, SS Aur and V442 Cen,
obtained as part of a Cycles 11 and 12 Snapshot Survey \citep{gan03}.
Since the success of our analysis rests heavily on having the best
possible distance estimates for each system, we examine distance estimates
in some detail. For all five systems, the distances, as in \citet{urb06},
were determined from \citet{war95} and \citet{har04} M$_{v(max)}$ versus
P$_{orb}$ relations, calibrated with trigonometric parallaxes. These
relations are affected by the inclination angle. Despite the large
uncertainties in the inclinations for all but eclipsing CVs, we examine this inclination
correction to our five systems in the next section.

The \citet{war95} relation is expressed as \[ M_{v(max)} =5.74 -
0.259P_{\rm orb} ({\rm hr}), \] and the \citet{har04} relation is \[
M_{v(max)} =5.92 - 0.383 P_{\rm orb} ({\rm hr}). \] The resulting
distances are compared with estimates based on other distance techniques
when available. A summary of the known properties of these systems is
given in the subsections below.

\subsection{System Parameters}

The distances we have estimated based upon the \citet{war95} and \citet{har04} 
relations are subject to a correction imposed by the system inclination since the apparent
luminosity of a disk depends sensitively on its inclination.  
This correction from apparent absolute magnitude to absolute magnitude 
(see Paczynski \& Schwarzenberg-Czerny 1980; Warner 1995) is given by
\begin{equation} 
\Delta M_v(i)  = -2.50 \log{ \left[ \left(1 + 
\frac{3}{2} \cos(i) \right) \cos(i) \right] } 
\end{equation} 
Note that $\Delta M_v(i) = 0$ for $i = 56.7$ degrees.

Three of the ''best'' inclinations for our systems are 60 degrees which is
close to the 56.7 degree value giving a correction of essentially zero for
three of the 5 systems. For BD Pav with $i = 71$ degrees, the inclination
correction to the M$_{v}$ is only +0.79 mag. For SS Aur ($i = 38$ degrees)
however, the inclination correction to the M$_{v}$ is -0.59. With the
inclination correction of delta(M$_{v}) = -0.59$ for $i = 38$ degrees, then
the Warner relation yields d = 200 pc which is virtually identical to the 
trigonometric parallax distance of 201 pc. With no inclination correction, then the
Warner relation yields d = 151 pc while the Harrison et al.(2004) relation
yields 178 pc.
 
For TU Men, the orbital period was determined by \citet{sto84} and
confirmed by \citet{men95}. Its orbital period actually places it in the
period gap. This dwarf nova is the longest period SU UMa system. It has
normal outbursts lasting a day and separated by 37 days of quiescence, and
superoutbursts lasting 4 to 20 days, repeating every 194 days.
Unfortunately, one dynamical solution is inconsistent with the tidal disk
instability theory while another K value is consistent with the theory of
superhump formation but may be contaminated by low velocity emission
features. \citet{men95}'s study yields a range of inclination angle $i =
44 - 52^{o}$ while the mass estimates range from M$_{wd} = 0.785$ to $1.06
M_{\sun}$. This object is one of the rare SU UMa systems in the period
gap. IUE archival spectra of TU Men were used to assess the reddening from
the 2175A ''bump'' (the visual inspection method described in
\citet{ver87}). We estimated E(B-V) = $0.07\pm 0.02$.  The STIS spectrum
was de-reddened by this amount in preparation for the fitting. For the
distance of TU Men, the \citet{war95} and \citet{har04} relations yielded
210 pc and 225 pc respectively. We estimate a distance d = 210 pc from the
donor tracks in \citet{kni06} for a predicted spectral type M4, the
K(2MASS) = 13.85 and a predicted M$_K$ = 7.2, if the secondary is
contributing 100\% of the flux in the K-band. In the present analysis we
adopted d = 210 pc.

BD Pav is one of the better studied systems among our snapshot targets.
The spectral type of the secondary, orbital period, secondary mass, mass
ratio, inclination angle and individual masses were determined by
\citet{axe88}. Further analysis was carried out by \citet{fri90a,fri90b}.
We estimated d = 420 pc, using the track in \citet{kni06} for K(2MASS) = 12.90, 
spectral type M3, and $M_K$ (predicted) = 6.0. We adopted d = 500 pc in the absence of 
any reddening information, which is modestly larger than the \citet{war95} value of 
375 pc and \citet{har04} relation value of 443 pc.

SS Aur has an orbital period $P_{orb}$ = 0.1828 days \citep{shahar86}.  
The mass of its white dwarf has been estimated to be $M_{wd} =
1.08\pm0.40$ $M_{\odot}$, while its secondary was found to have a mass
$M_{2} = 0.39\pm 0.02$ $M_{\odot}$, and a system inclination $i =
38^{o}\pm16^{o}$ \citep{sha83}. SS Aur has a distance of 201 pc from an
{\it{HST}} FGS parallax measurement of 4.97 mas \citep{har99} and a
reddening value $E(B-V)=0.08$ \citep{ver87,lad91,bru94}. It is useful to
compare what the best value of the distance to SS Aur would be in the
absence of the trigonometric parallax. The method of \citet{kni06} places
a strong lower limit on the distance by assuming a 100\% donor
contribution in K-band. For SS Aur, the apparent K-mag is 12.02, yielding
a predicted donor absolute K-magnitude of 5.74 and a distance of 180 pc.
The Warner and Harrison et al. relations yield 151 pc and 178 pc, both
values below \citet{kni06}.

For TT Crt, the earliest study was conducted by \citet{szk92} who
determined an orbital period of 0.30 days, the spectral type of the
secondary, mass ratio, inclination, and mass of the primary. More
recently, \citet{tho04} carried out an extensive radial velocity study of
TT Crt and found its orbital period to be P$_{orb} = 0.26835$ days. The
\citet{tho04} study suggests the values M$_{wd} = 1$ M$_{\sun}$ and $i =
60^{o}$. For TT Crt we took d = 525 pc which is the value given by
\citet{war95}'s relation but is closer than estimates from the secondary
spectral type (K$5\pm1$) which gives 760 pc or the \citet{har04} relation
(708 pc).

V442 Cen is a dwarf nova but little else is known from optical studies.
\citet{mar84} found the orbital period to be 0.46 days. \citet{mun98}
detected a weak H-$\alpha$ emission core which is essentially the only
evidence that this system has a disk and is a dwarf nova. The mass of the
WD and the inclination are unknown. For the distance of V442 Cen, we
adopted d = 800 pc which is roughly midway between the values given by the
\citet{war95} relation (637 pc) and the \citet{har04} relation (1136 pc).
Any distance from these relations must be viewed with caution since the
orbital period of V442 Cen itself is uncertain. The reddening of V442 Cen
($E(B-V)=0.15$) is found only in the work of \citet{war76}, who also
assessed the reddening of 6 additional DNe in quiescence (Z Cam, V436 Cen,
WW Cet, SS Cyg, U Gem and VW Hyi). However, more recent estimates of the
$E(B-V)$ value for Z Cam, V436 Cen, WW Cet, U Gem and VW Hyi are all at
least a factor of two smaller than the $E(B-V)$ values listed in
\citet{war76} for these objects. We therefore regarded the value 0.15 as
an upper limit for the reddening of V442 Cen and we adopted a value of
0.10 for our model fits.

For all five systems, we followed the same procedure for estimating
distances as in \citet{urb06} where the \citet{war95} and \citet{har04}
M$_{v(max)}$ versus P$_{orb}$ relations, calibrated with trigonometric
parallaxes, were used. These distances will be discussed again in Section 4
below. The reddening values are from \citet{ver87},
\citet{lad91} and \citet{bru94}. In Table 1 we summarize the observed
properties of the five dwarf novae; - these are also the values we adopted
in our modeling (see also \citep{rit03}). The columns are as follows: (1)
system name; (2) dwarf nova subclass with UG denoting a U Gem-type system,
ZC denoting a Z Cam-type system; (3) orbital period in days; (4) the
recurrence time of dwarf nova outbursts in days;  (5) the apparent
magnitude at minimum (quiescence); (6)  the apparent magnitude in
outburst; (7) secondary spectral type; (8) orbital inclination in degrees;
(9) white dwarf mass in solar masses;  (10) adopted reddening value and ;
(11) adopted distance in parsecs.

\clearpage

\setlength{\hoffset}{-15mm}
\begin{deluxetable}{lllllllllll}
\tabletypesize{\small}
\tablecaption{Dwarf Nova Parameters}
\tablewidth{0pc}
\tablehead{
\colhead{System}&
\colhead{Subclass}&
\colhead{$ P_{\rm orb}$}&
\colhead{$t_{\rm rec}$}&
\colhead{V$_{\rm min}$}&
\colhead{V$_{\rm max}$}&
\colhead{Sec.Sp.Typ.}&
\colhead{$i$}&
\colhead{$M_{wd}$}&
\colhead{$E(B-V)$}&
\colhead{Distance} \\
\colhead{}  &
\colhead{}  &
\colhead{(d)}  &
\colhead{(d)}  &
\colhead{}  &
\colhead{}  &
\colhead{}  &
\colhead{(deg)}  &
\colhead{($M_{\odot}$)}  &
\colhead{}  &
\colhead{(pc)}
 }
\startdata
TU Men & UG & 0.1172  & 37  & 18.5 & 11.6 & M4V        & 60 & 0.8 & 0.07 & 210 \\
BD Pav & UG & 0.179301&     & 15.4 & 12.5 & M3V        & 71 & 1.15& 0.0 & 500 \\
SS Aur & UG & 0.1828  & 43  & 14.7 & 10.5 & M1V        & 38 & 1.1 & 0.08& 201 \\
TT Crt & UG?& 0.26835 &     & 15.9 & 12.7 & K5V$\pm0.75$& 60 & 1.0 & 0.0 & 525 \\
V442 Cen& UG& 0.46    &14-39& 16.5 & 11.9 & G4V        & 60 & 0.8 & 0.10& 800 \\
\enddata
\end{deluxetable}

\clearpage

\setlength{\hoffset}{00mm}

\section{Observations and Data Reduction}

FUV spectroscopy of the five dwarf novae was obtained with {\it{HST}}/STIS
during {\it{HST}} Cycle 11. The data were obtained using the G140L grating
and the $52^{\prime\prime} \times 0.2^{\prime\prime}$ aperture, providing
a spectral resolution of R$\sim 1000$ over the wavelength range 1140-1720
\AA. Since the total time involved in each snapshot observation was short
($< 35$min), the observations were made in the ACCUM mode in order to
minimize the instrument overheads. This resulted in exposure times of 600
to 900 seconds.  Each snapshot observation resulted in a single time
averaged spectrum of each dwarf nova. All of the spectral data were
processed with IRAF using CALSTIS V2.13b.During target acquisition,
{\it{HST}} points at the nominal target coordinates and takes a
$5^{\prime\prime} \times 5^{\prime\prime}$ CCD image with an exposure time
of a few seconds. Subsequently, a small slew is performed that centers the
target in the acquisition box, and a second CCD image is taken. The
acquisition images for these observations were obtained using the F28x50LP
long-pass filter, which has a central wavelength of 7228.5 and a
full-width at half maximum (FWHM) of 2721.6\AA\ \citep{ara05}. We
calculated the F28x50LP magnitudes as in \citet{ara05}.

The instrumental setup and exposure details of the {\it{HST}} STIS spectra
of TU Men, BD Pav, SS Aur, TT Crt, and V442 Cen are provided in the
observing log given in Table 2, the entries are by column: (1)  the
target, (2) Data ID, (3) Instrument Config/Mode, (4) Grating, (5) Date 
of Observation (yyyy-mm-dd), (6) Time of observation, (7) Exposure time 
(s), and (8) F28x50LP magnitude.

\clearpage
\setlength{\hoffset}{-23mm}
\begin{deluxetable}{lccccccc}
\tablecaption{{\it{HST}} Observations}
\tablewidth{0pc}
\tablehead{
\colhead{Target}              &
\colhead{Data ID}             &
\colhead{Config/Mode}            &
\colhead{Grating} &
\colhead{Date of Observation} &
\colhead{Time of Observation}  &
\colhead{t$_{exp}$(s)}&
\colhead{Mag}           

}
\startdata
TU Men  & O6L156010 &  STIS/FUV-MAMA/ACCUM & G140L &   2003-01-04 & 01h:47m:00s &  900&16.4  \\
BD Pav    &  O6LI0C010  & STIS/FUV-MAMA/ACCUM & G140L &   2003-02-09 & 13h:48m:00s &  600&14.6  \\
SS Aur    &  O6LI0F010  & STIS/FUV-MAMA/ACCUM & G140L &  2003-03-20    & 11h:49m:00s &   600&15.5  \\
TT Crt  &  O6LI1K010 & STIS/FUV-MAMA/ACCUM & G140L &   2003-02-12 & 05h:12m:00s &  700&15.1  \\
V442 Cen  &  O6LI1V010 &  STIS/FUV-MAMA/ACCUM & G140L &   2002-02-09 & 09h:07m:00s &  600 &15.4 \\
\enddata
\end{deluxetable}

\clearpage

\setlength{\hoffset}{00mm}

The spectrum of the dwarf nova TU Men reveals a broad Ly$\alpha$
absorption with both wings well-defined, C\,{\sc iii} 1175 absorption,
possible weak N\,{\sc v} 1240 absorption, Si\,{\sc ii} 1260, 1265, a
possible C\,{\sc i} absorption at 1280\AA, O\,{\sc i} + Si\,{\sc iii}
(1300\AA), C\,{\sc ii} 1335, Si{\sc iv} 1393, 1402 in weak emission and
very strong C\,{\sc iv} 1548, 1550 emission.
The strong C IV emission has a velocity width suggesting its formation in the 
optically thin accretion disk present during quiescence. It forms at a lower 
temperature than N V while the absence of Si IV may be an abundance effect. 
The spectrum of TU Men is identified with the hot accreting white dwarf.

For BD Pav, a strong continuum and Ly$\alpha$ absorption wings are seen.
The longward wing is flanked by a strong N\,{\sc v} 1240 emission feature.
There is strong, presumably photospheric, absorption due to Si\,{\sc ii}
1260, 1265, Si\,{\sc iii} + O\,{\sc i} 1300, weak, narrow C\,{\sc ii} 1335
absorption, weak Si\,{\sc iv} 1393, 1402 absorption/emission and narrow
Si\,{\sc ii} 1526, 1533 absorption. Based upon the strength of N\,{\sc v}
and the absence of C\,{\sc iv} 1550, it appears that BD Pav is another
cataclysmic variable that reveals a N/C composition anomaly \citep{gan03}.

In over 80\% of the cataclysmic variables, the relative strengths of NV
and C IV are commensurate with what one expects for a solar composition
accretion disk.  This abundance anomaly wherein the C IV emission feature
is very weak or absent while the N V is very strong and dominant appears
in less than 20\% of the CVs. Yet the continuum slopes and Lyman Alpha
profiles of the accretion disks do not indicate an extraordinarily hot
accretion disk or white dwarf. It is compelling therefore that this is an
abundance effect signalling that the disk and hence the accreting material
has undergone CNO processing.

SS Aur's STIS spectrum reveals a strong Ly$\alpha$ absorption and numerous
somewhat weak metal absorption lines of C\,{\sc iii} 1175, Si\,{\sc ii}
1260, 1265, Si\,{\sc iii} 1300, C\,{\sc ii} 1335, Si\,{\sc iv} 1393, 1402,
unidentified absorption at 1500\AA\ which could be Si\,{\sc iii}, strong
emission at C\,{\sc iv} 1550 and a hint of emission at N\,{\sc v} (1238,
1242). There appears to be weak emission wings flanking the central
absorption at C\,{\sc ii} 1335 and Si\,{\sc iv} 1393, 1402.

TT Crt's spectrum is characterized by a rich absorption line spectrum with
C\,{\sc iii} 1175, two Si features between 1180 and 1190, Si\,{\sc ii}
1260, 1265, 1526, 1533, N\,{\sc v} 1238.8 \& 1242.8, S\,{\sc i} 1247,
S\,{\sc ii} 1250, Si\,{\sc iii} 1300, C\,{\sc ii} 1335, Si\,{\sc iv} 1393,
1402, Si\,{\sc ii} 1530, C\,{\sc iv} 1550 and a possible He\,{\sc ii} 1640
absorption. The continuum rises into the far UV indicating a hot white
dwarf as the most likely source of the continuum and the absorption lines.
In addition, the weak absorption feature around 1350\AA\ could be a blend
of C\,{\sc i}+O\,{\sc i}+Si\,{\sc ii}, and the feature around 1370 could
be due to either Ni\,{\sc ii}, or P\,{\sc i}, or possibly O\,{\sc v}1371
absorption. The mix of absorption features from different ionization
stages suggests line formation in two different temperature regions. The
N\,{\sc v} doublet (1240), C\,{\sc iv} (1550) and He\,{\sc ii} (1640)
clearly must form at temperatures well above the temperature of the WD
expected from its continuum slope and Ly$\alpha$ profile while the low
ionization features would be nearly absent for $T_{eff}> 25,000$K.

For V442 Cen, a continuum rising into the blue is seen. On the longward
wing of Ly$\alpha$ there is what appears to be N\,{\sc v} 1238\AA, 1242,
emission and a prominent C\,{\sc iv} emission feature. There may be
absorption at C\,{\sc iii} 1175 but the data is too noisy for the
identification to be certain. However, there is a probable weak emission
feature at Si\,{\sc iii} + O\,{\sc i} (1300). There is also a weak
unidentified absorption line near 1500\AA (which we also suspect to be
Si\,{\sc iii}).

The spectra of the five dwarf novae all exhibit a very sharp Ly$\alpha$
emission line (of a few $10^{-15}$ergs$~$s$^{-1}$cm$^{-2}$\AA$^{-1}$
centered around 1218\AA ) due to air glow, as well as a very sharp
Ly$\alpha$ absorption feature due to interstellar or possibly circumbinary
atomic hydrogen centered at 1215.7\AA. The exact width of the bottom of
the Ly$\alpha$ absorption feature cannot accurately be determined because
of the airglow and the low flux level ($\sim
10^{-15}$ergs$~$s$^{-1}$cm$^{-2}$\AA$^{-1}$) for some of the objects (e.g.
BD Pav). It appears, however, that this ISM absorption feature is wider in
the spectrum of V442 Cen in agreement with its higher reddening value. Of
all the targets, only SS Aur and V442 Cen have a known reddening. For TU
Men we assess the reddening here in this work, and the other two targets
could also have a non-zero reddening.  Unfortunately, the FUV data needed
to assess $E(B-V)$ for the other two systems is sparse.

\section{Multi-Component Synthetic Spectral Models}

Our data analysis and modeling for the {\it{HST}} STIS range involve the
full suite of multi-component (accretion disk, white dwarf photosphere,
accretion belt, absorbing curtain, etc.)  synthetic spectral codes, which
we have utilized in the past in our spectral fitting of {\it{HST}},
{\it{FUSE}}, {\it{HUT}}, {\it{EUVE}}, and {\it{IUE}} data. Based upon our
expectation that the accreting white dwarf is an important source of FUV
flux in these systems during quiescence, we carried out a high gravity
photosphere synthetic spectral analysis first. Model atmospheres were 
computed using the code TLUSTY\footnote{version 200, http://nova.astro.umd.edu}
(Hubeny 1988; Hubeny \& Lanz 1995) and the synthetic spectra using the code 
SYNSPEC version 48 (Hubeny \& Lanz 1995). The details of our 
$\chi^{2}_{\nu}$ (per degree of freedom) minimization fitting procedures are 
discussed in detail in \citet{sio95b} and will not be repeated here.
 
To estimate physical parameters, it is common to take the white dwarf
photospheric temperature T$_{eff}$, surface gravity log $g$, and projected
rotational velocity $v_{rot} \sin{i}$ as free parameters. However, in the
FUV region, gravities derived from fitting the profile of Lyman Alpha must
be viewed with caution since lower n transitions of H are the least
sensitive to surface gravity.  

In this work, depending upon our confidence in the distance of a given 
system, we either adopted the preferred estimate of the distance to each 
system in Table 1 by fixing it at a constant value in the fitting or found 
the scale factor-derived distance from the best fit that was closest the 
preferred distance from Table 1. Then, the WD model fits yield a white 
dwarf radius for the system, which can be converted to a mass from a 
mass-radius relation. For the mass-radius relation, we used the 
evolutionary model grid of Matt Wood (1995) for C-O cores and 
hydrogen envelopes. In the absence of higher quality data and to limit the 
complexity of the analysis, we fixed the abundances at solar. Our choice 
to fix the abundances at solar affects somewhat our determination of the 
rotational velocities, because the effects of rotational velocity and 
abundances are intertwined.  However, increasing the number of fitted 
parameters is not justified here, given the limited statistical quality of 
the data (and the fact that a number of the systems have composite 
spectra).

The models are normalized to 1 solar radius and 1 kiloparsec such that the 
distance of a source is computed from $d = 
1000(pc)*(R_{wd}/R_{\sun})/\sqrt{S}$, or equivalently the scale factor $S 
= \left( \frac{R_{wd}}{R_{\odot}} \right)2 \left( \frac{d}{kpc} 
\right)^{-2}$, is the factor by which the theoretical flux (integrated 
over the STIS wavelength range) has to be multiplied to equal the observed 
(integrated) flux.

Unless otherwise specified, the grid of white dwarf models extended over
the following range of parameters: log $g = 7.0, 7.5, 8.0, 8.5, 9.0$;
T$_{eff}/1000$ (K) = 11, 12, ..., 75; abundances fixed at solar; and
$v_{rot} \sin{i}$ (km~s$^{-1}$) = 100, 200, 400, 600, 800. The accretion
disk models were adopted from the optically thick, steady state model grid
of \citet{wad98}.

For each dwarf nova, we masked out all of the obvious emission features
and artifacts in the {\it{HST}} STIS spectrum of each object. In Table 3,
we indicate these masked regions where strong emission features,
artifacts, or negative fluxes occur.

\clearpage
\begin{deluxetable}{lc}
\tablecaption{Masked Spectral Regions}
\tablehead{
\colhead{Target}&\colhead{Emission Features }
}
\startdata
TU Men    &    1380-1403, 1535-1565  \\
BD Pav    &    1208-1219, 1237-1243, 1385-1400  \\
SS Aur      &     1540-1560 \\
TT Crt    &    $<1180$, 1195-1205, 1235-1245, 1530-1560 \\
V442 Cen    &    1237-1245, 1540-1560, 1650-1662   \\
\enddata
\end{deluxetable}
\clearpage

\section{Spectral Fitting Results}

For dwarf nova quiescence, the most obvious first choice for modeling the
FUV flux distribution and absorption lines is a single temperature white
dwarf since according to theory of dwarf novae outbursts the inner disk
should be optically thin or absent during quiescence (Cannizzo 1998) and the
accretion-heated white dwarf should be the dominant source of flux. While
this is the case in many quiescent dwarf novae, there are numerous
exceptions where a second component of FUV flux is required in order to
account for the observed shape of the flux distribution. Since the exact
physical nature of this additional flux source is presently unknown,
various experimental representations have been tried including simple
black bodies, power laws, optically thick accretion disks (in the absence
of optically thin disk models) and two-temperature white dwarf models in
order to simulate a hotter equatorial region as well as a cooler
photosphere at higher latitudes. For dwarf novae above the period gap,
mass transfer rates are expected to be higher than for systems below the
period gap. Thus, accretion rates during quiescence may also be higher
than for systems below the gap (Warner 1995, Cannizzo 1998). Noting that 
optically thick disks in many cases resemble the continuum shapes and 
even reproduce absorption line profiles of some long period dwarf novae 
during quiescence, we have used these disk models as proxies to represent
the second components of FUV flux of the objects in this work.

For each dwarf nova, the synthetic spectral fitting was carried out with
single model components or combinations of model components in the
following order: a single temperature white dwarf model alone, an
accretion disk model alone, a combination white dwarf model plus accretion
disk model, and a two-temperature white dwarf model. The results of the
model fitting are given in Table 4 where the entries are by column: (1)
System Name;  (2) Model type; (3) white dwarf mass, M$_{wd}$; (4) WD
effective surface temperature T$_{eff}$; (5) WD gravity Log g; (6) WD
projected rotational velocity $v_{rot}\sin{i}$; (7) 2nd component
temperature; (8) disk inclination i; (9) mass accretion rate in
$M_{\odot}$yr$^{-1}$; and (10) minimum $\chi^{2}_{\nu}$ achieved.

\clearpage
\begin{deluxetable}{lccccccccc}
\tabletypesize{\small}
\tablecaption{Spectral Fitting Results}
\tablewidth{0pc}
\tablehead{
\colhead{System}   &
\colhead{Model}             &
\colhead{M$_{wd}$}            &
\colhead{T$_{eff}$} &
\colhead{Log g}           &
\colhead{$v_{rot}\sin{i}$} &
\colhead{T2}  &
\colhead{i}   &
\colhead{$\dot{M}$} &
\colhead{$\chi^{2}_{\nu}$}  \\
\colhead{} &
\colhead{} &
\colhead{($M_{\odot}$)} &
\colhead{(K)} &
\colhead{(cgs)} &
\colhead{(km$~$s$^{-1}$)} &
\colhead{(K)} &
\colhead{(deg)} &
\colhead{($M_{\odot}~yr^{-1}$)} &
\colhead{}
}
\startdata

TU Men   &    WD   & 0.8 & 28,000 & 8.3 & 400: & --     & -- &  --                 &  1.81  \\
         &   Disk  & 0.8 &  --    & --  &  -- & --     & 60 & $5 \times 10^{-11}$ &  6.379  \\
         & WD+Disk & 0.6 & 20,000 & 8.0 &  -- & --     & 60 & $3 \times 10^{-12}$ &  3.546  \\
         &  2-T WD & 0.6 & 26,000 & 8.0 & 600: & 40,000 & -- & --                  &  1.79  \\
BD Pav   &  WD     & 1.2 & 27,000 & 9.0 & 600: & --     & -- & --                  & 13.875  \\
         & Disk    & 1.2 &   --   & --  & --  & --     & 75 & $6 \times 10^{-11}$ & 10.026  \\
         & WD+Disk & 1.2 & 21,000 & 9.0 & 600: & --     & 75 & $ 3\times 10^{-11}$ &  7.671  \\
         & 2-T WD  & --  &  --    & --  &  -- & --     & -- &   --                & 13.809  \\
SS Aur   &   WD    & 1.1 & 34,000 & 8.8 & 400: & --     & -- & --                  &  2.15   \\
         &   Disk  & 1.0 & --     & --  & --  & --     & 60 & $ 3\times 10^{-10.5}$ & 10.72 \\
         & WD+Disk & 1.0 & 27,000 & 8.7 & 400: &  --    & 60 & $ 3\times 10^{-10.5}$ &  5.68 \\
         & 2-T WD  & 1.1 & 30,000 & 8.8 & 600: & 40,000 & -- &      --             &  2.09  \\
TT Crt   &  WD     & 0.9 & 29,000 & 8.5 & 200: &  --    & -- &    --               &  3.80  \\
         & Disk    & 1.0 &  --    & --  & --  &  --    & 60 & $8 \times 10^{-11}$ &  7.818  \\
         & WD+Disk & 1.0 & 30,000 & 8.7 & 200: &  --    & 60 & $3 \times 10^{-11}$ &  4.711  \\
         &  2-T WD & 0.9 & 31,000 & 8.5 & 200: & 45,000 & -- & --                  &  3.77  \\
V442 Cen &  WD     & 0.8 & 47,000 & 8.3 & 200: & --     & -- &   --                &  2.95   \\
         & Disk    & 0.8 &  --    & --  & --  &  --    & 60 &  --                 &  8.7    \\
         & WD+Disk & 0.8 & 41,000 & 8.3 & 600: &  --    & 60 & $6 \times 10^{-10}$ &  5.538  \\
         &  2-T WD & 0.8 & 40,000 & 8.3 & 200:  & 50,000 & -- &   --                &  2.67  \\
\enddata
\end{deluxetable}

\clearpage

We note that in table 4, there are cases of WD plus accretion disk fits
which have larger $\chi^{2}$ values than the WD fit alone. Intuitively,
one expects the value for the combined fit to never be larger than the
white dwarf-only fit. These cases result from our disk model grid (the
Wade and Hubeny grid) not being sufficiently broad in parameter space at
the low Mdot end of the grid.  Since the grid does not extend to very low
values of the accretion rate, the fit can never force the disk component
to become essentially zero, i.e. to reach the $\chi^{2}$ value of the
WD-only fit.

As seen in figure 1, the best fitting models to TU Men, the SU UMa system
with the longest orbital period, proved to be a single temperature white
dwarf fit.  We note that our comparison of the model line profiles of
metals to the STIS data revealed a redshift of about 0.7 \AA. This shift
corresponds to a velocity range between 150 and 200km/sec for FUV
wavelengths of 1500\AA\ down to 1000\AA, respectively. When the model was
shifted by 0.7 \AA\ to match the STIS data, the $\chi^2_{\nu}$ of the best
fit was reduced to 1.81, the best fitting single temperature WD to the
dereddened (E(B-V) = 0.07) corresponded to T = 28,000K (for an assumed Log
g = 8.3), with the model fits to the narrow metal lines giving
$v_{rot}\sin{i} = $400 km/s, and a scale factor-derived distance d = 288
pc, as shown in figure 1. This distance is considerably larger than our
favored distance of 210pc in Table 1. When we assume M=1.0 (log g =8.64)
and obtained a new temperature, this led to d = 214 pc. A word of caution
is advisable here on how the uncertainty of the distance to TU Men and the
other systems translates to an error in the derived temperature. For each
given combination of T and log g, the scale factor yields a different
distance. At a given log g, we can estimate the error in distance to the
uncertainty in temperature. For very high temperature objects, the flux
goes like T/d**2 so in the high temperature limit one has (delta d)/d =
(0.5 delta T)/(T). Thus, if we have an uncertainty in the distance of,
say, 10\% closer or further, then the temperature error could be as high
as 20\%. This underscores how critical it is to have accurate distances
not only to have the correct white dwarf temperatures but also to be able
to properly disentangle the various FUV emitting components (e.g. disk,
WD, belt etc.).

The STIS spectrum could not fit successfully with an optically thick disk
structure or even improved with a combination of a WD and a disk. A
two-temperature white dwarf fit also failed to improve the quality of the
fit. In all but the single temperature white dwarf case, the model
Ly$\alpha$ profiles were too broad if the continuum slope matched the
data.  We also tried two-temperature white dwarfs (WD + accretion belt).
The best 2-T fit had T$_{wd}= 26,000$K, T$_{2}= 40,000$K, with a $\chi^{2}
= 2.11$. The area of the belt is 7\% and contributes about 40\% of the
total flux. If T$_{wd}$ was decreased further, a larger $\chi^{2}$
resulted.

For BD Pav, the observed continuum slope and flux level together cannot be
matched by a single component fit or by combination fits. The most
successful fit is shown in figure 2 and is still far from satisfactory. It
is a combination of a massive WD (log g = 9) with T$_{eff} = 21,000$K and
an optically thick disk component with an inclination $i = 75^{o}$. The
disk component provides the longward continuum flux to match the observed
flux level and slope but there is a model flux level excess at wavelengths
below Ly$\alpha$ relative to the data. As seen in figure 2, the two
components contribute, on average, 77\% of the flux and 23\% of the flux
for the disk and white dwarf, respectively. In view of the relatively poor
quality of the STIS spectrum of this object, we have eliminated it from
any further discussion in this paper.

For SS Aur, the dereddened STIS spectrum is best fit by a white dwarf
alone as shown in figure 3.  For a distance of 200 pc and a solar
abundance WD, the best WD model has a WD mass of $1.1 M_{\odot}$ (log
g=8.8), $T_{eff} = 34,000 \pm 2000$K with a projected rotational velocity
of $v_{rot}\sin{i}$ = 400 km/sec. In an {\it{IUE}} study of SS Aur during
quiescence, \citet{lak01} found initial indications that the FUV spectrum
was dominated by a hot photosphere with T$_{eff} = 30,000$K, log $g$ =
8.0. The parallax together with the scale factor S $= 1.12\times 10^{-3}$
from their best fit, yielded a radius for the emitting source of
$4.68\times 10^{8}$ cm corresponding to a white dwarf. \citet{sio04} found
that the favored solution to a {\it{FUSE}} spectrum of SS Aur in
quiescence, was a white dwarf of 27,000K plus a 48,000K accretion belt. We
note that in figure 3, the fit to the bottom of the Ly$\alpha$ region is
not perfect as the ISM atomic hydrogen drives the flux to zero there.
Since the fit is otherwise accurate and in good agreement with the
distance and mass, we decided to exclude ISM atomic hydrogen opacity in
the modeling. A disk alone or a combination of a disk + WD models were
inconsistent with either the 200pc distance and/or the expected low mass
accretion rate $\dot{M}$ during quiescence. The disk models in our grid
with inclinations closest to the empirically-derived value of 38 degrees
had values of 41 degrees. To our surprise, the best-fitting disk-only and
WD + disk models had inclinations of 60 degrees.
  
In the best fitting WD + disk model, the WD accounts for 60\% of the FUV
flux while the disk accounts for 40\%. Most of these models had a
Ly$\alpha$ profile too broad and/or the slope of the continuum could not
be matched. The white dwarf plus accretion belt fit likewise failed to
improve the fits to the STIS data.  The best-fitting 2-T models had
T$_{wd}=30,000$K, T$_{2}=40,000$K, with $\chi^{2} =2.53$ for d = 193pc and
a belt area of 0.15. As the area of the belt decreased we obtained lower
$\chi^{2}$ for an area below 8\% but then the distance drops below the
parallax distance of 201 pc. As the area of the 2nd component was
increased up to 20\%, then the distance increases to 200pc, but the
$\chi^{2}$ becomes larger. We therefore disregarded these models
altogether.

TT Crt offered the richest metallic line spectrum for fitting with solar
composition models and, with its Ly$\alpha$ absorption depth reaching
close to zero, a single temperature WD would have seemed the strongest
candidate source. Indeed, a single temperature WD does provide an
excellent fit to Ly$\alpha$ and fits the widths of the metallic absorption
lines, but the match to the continuum slope is less than satisfactory and
yields a pronounced model flux deficit relative to the data shortward of
$\sim 1520$\AA. If E(B-V) = 0.0, the best single temperature WD fit (with
log g = 8.3) is $T_{eff} = 29,000$K, $v_{rot} \sin{i}= 200 \pm 100 $km/s
with a scale factor-derived distance of 506 pc. This value of distance is
quite close to our adopted distance in Table 1. The single temperature WD
fit is displayed in figure 4a. However, in view of the relatively large
distance we favored, it was prudent to examine the effect on the fits if
reddening is assumed. For this exercise, we used both the STIS data in
quiescence and IUE data of TT Crt in outburst. We assumed different values
of E(B-V) (0.00, 0.05, 0.10, 0.15)  for both the STIS observations and the
IUE observations. We modeled the STIS data with a WD only and the IUE
outburst data with a disk only but the best result (for which IUE and STIS
results are consistent) is for E(B-V)=0.15 which yields a distance of
about 220 pc and $T_{eff} = 30,000$K for a $1.0 M_{\odot}$ white dwarf.
This distance is well below our adopted value from Table 1.
  
For lower values of E(B-V), the fits between IUE and STIS are discordant,
e.g. E(B-V)= 0 leads to a distance of 375pc for STIS and 100pc for IUE
(M=1.0). In fact, the reddening constrains the distance to be less than
500 pc because the value 500pc corresponds to E(B-V) = 0.0 for a white
dwarf mass of $0.8 M_{\odot}$ or greater. To get a larger distance, one
would have to assume a much smaller mass which is inconsistent with
previous work. In any case, a large distance would be consistent with a
larger E(B-V) but as E(B-V) increases, the resulting distance decreases
(as the flux increases).

An optically thick disk structure alone does a poor job of matching the
data because it utterly fails to fit any of the sharp metallic lines,
mismatches the continuum slope, and has a marked flux deficit relative to
the data. The situation is significantly improved with combinations of a
WD plus an optically thick disk and a two-temperature WD. The 2-T WD model
with $T_{eff} = 31,000$K and 45,000K matches the continuum slope and sharp
metallic lines as well as Ly$\alpha$ but there remains a flux deficit
between roughly 1600 \AA\ and 1250 \AA. The combination of a WD with
$T_{eff} = 30,000$K and an optically thick disk accomplishes everything
the 2-T WD does and reveals less of a flux deficit though a deficit of
model fluxes relative to the data is still evident shortward of 1400 \AA.
In figure 4b, the best-fitting WD + disk combination is shown for
comparison.

V442 Cen presented a number of challenges. Since it has a large reddening
and distance, its very likely that the Ly$\alpha$ feature is mainly due to
the ISM atomic hydrogen opacity. We therefore decided to include in our
modeling the ISM atomic hydrogen by using an ISM model with an atomic
hydrogen column density N(H\,{\sc i})=$5 \times 10^{20}$cm$^{-2}$, a
turbulent velocity of 40 km/s, and a temperature of 170K. Such a column
density is consistent with the reddening value we adopted (e.g.  
\citet{boh78}). These more realistic WD fits to the {\it{HST}} STIS
spectrum of V442 Cen with E(B-V) = 0.10 yielded a best-fitting model
consisting of a 47,000K white dwarf with Log $g= 8.3$ for a distance of
only 328 pc. This distance is well below even the value of 637 pc obtained
from the Warner relation. This model, seen in figure 5, had a
$\chi^2_{\nu}=2.95$. Using the ISM model, we also tried a disk alone and
disk+WD models but they did not lead to any improvement in the fit. The
disk alone models all had a very large $\chi^2_{\nu}$ and the least
$\chi^2_{\nu}$ that could be obtained with a disk model was 8.7, while the
disk+WD models had at best $\chi^2_{\nu}=5.5$. We also tried
two-temperature WD models. The best-fitting 2-T WD model had a best
$\chi^2_{\nu}=2.67$, comparable to, but not significantly better than, the
best single temperature WD model.

It is important to note that by our fixing the metal abundances at their
solar value rather than allowing for metal abundance variations in the synthetic
spectral fitting, systematic errors are introduced that carry through 
to other parameters. There are two kinds of errors, those
that drive the $\chi^{2}$ up and those that increase the errors
in all the other parameters. The error in $v_{rot} \sin{i}$ affects the abundances and 
$\chi^{2}$ but not the other parameters, while the error in the abundances affects 
$v_{rot} \sin{i}$ and $\chi^{2}$ but not the other parameters. The temperature error affects $\chi^{2}$, 
the distance and gravity while the error in log g affects $\chi^{2}$, the distance 
and temperature. The error in distance affects $\chi^{2}$, log g and temperature.
Of these parameters, the distance, mass and temperature are more fundamental than
the abundances and rotation since they affect the whole continuum
while $v_{rot} \sin{i}$ and abundances affect only the absorption line fits.  

Finally, the reader should be aware that none of the fits that we have
obtained for the five systems are formally acceptable fits from a strict
mathematical point of view and thus one must use them with caution. 
Indeed, from a mathematical perspective, the probability that we have the
correct model with a $\chi^2_{\nu}=2.0$ being our best value, is
essentially zero. Hence, from this standpoint it is very difficult to
argue that one model is really better than another.

\section{Conclusions}

The five long period dwarf nova systems analyzed here, together with
previously analyzed systems, now comprise a sample of 13 systems for which
we have estimates of the white dwarf surface temperatures and their
rotational velocities. This helps to remove the disparity between the
sample of CV WDs with known temperatures below the period gap and the
number above the gap. The white dwarf temperatures are typically uncertain
by ± 2000K while the rotational velocities must be regarded with extreme
caution since they are influenced by what one determines for the metal
abundances. Without a precision abundance analysis, it becomes difficult
to disentangle the rotational effects on the line profiles of metals from
abundance effects. Thus, it is unlikely that the velocities we quote in
this work are the true rotational velocities of the white dwarfs. All that
one can state with any confidence is that the rotational velocities
derived up to now for white dwarfs in dwarf novae are of order 10\% (or
less) of the typical Keplerian velocity at the white dwarf surface.

It is also important to point out that the model fitting we have carried out
is in some sense rather crude and overly simplistic. The components are
combinations of single or two-temperature white dwarfs and classical
disks. There are two important astrophysical processes that our modeling
must eventually incorporate. First, we have totally ignored the effects of
irradiation of the disk by an accretion-heated, hot white dwarf. The
inclusion of irradiation may in fact improve the fits. Second, line
profiles that arise, either in emission or in absorption (depending on the
observers line of sight with respect to the system), from circumsystem
material surrounding the white dwarf and accretion disk can be calculated
using CLOUDSPEC (Hubeny, unpublished). This routine computes the detailed 
line profiles arising from circumsystem material. Finally, the model fitting
was carried out with empirically-constrained values of the surface gravity.
For higher or lower values of the gravity, the best-fit temperatures would
increase or decrease by, of order, 2000 to 3000K. 

In Table 5, we list these systems with their white dwarf temperatures in
order of their orbital period in minutes beginning with the longest
period. The snapshot targets need {\it{FUSE}} to better constrain the
T$_{eff}$ but unfortunately the {\it{FUSE}} mission suffered a failure and 
had to be terminated.   The high inclination of some systems may hide much of the WD
photosphere. Also some systems with unknown reddening might have a large
$E(B-V)$ value and/or ISM atomic hydrogen absorption, therefore affecting
the estimate of the temperature. However, without better knowledge of the
reddening and the atomic hydrogen column density, and without {\it{FUSE}}
spectra it is impossible to find out how the Ly$\alpha$ profile of these
objects is affected by the ISM.
\clearpage
\begin{table}
\caption{Surface Temperatures of White Dwarfs in Dwarf Novae above the Period Gap}
\begin{tabular}{lccl}
\hline
\hline
System Name &  Period & $T_{eff}$ &     References   \\
        &   (min)  &  (Kelvin)    &                 \\ \hline                
BV Cen  &   878.6  &   40,000     &    \citet{sio07}       \\
V442 Cen &  662.4  &   47,000     &    This paper          \\
RU Peg  &   539.4  &   49,000     &    \citet{sio04}       \\                
Z Cam   &   417.4  &   57,000     &    \citet{har05}       \\                
SS Cyg  &   396.2  &   55,000     &    \citet{sio07b}       \\
TT Crt  &   386.4  &   30,000     &    This paper          \\
RX And  &   302.2  &   34,000     &    \citet{sio01}       \\
SS Aur  &   263.2  &   34,000     &    This paper          \\
U Gem   &   254.7  &   31,000     &    \citet{sio98}       \\
WW Ceti &   253.1  &   26,000     &    \citet{god06}       \\
UU Aql  &   230.4  &   27,000     &    \citet{sio07}       \\
TU Men  &   168.8  &   28,000     &    This paper          \\
\hline
\end{tabular}
\end{table}
\clearpage

A preliminary look at the distribution of dwarf nova white dwarf
temperatures above the period gap, suggests that for orbital periods,
P$_{orb}$, between $\sim 200$ minutes and minutes and $\sim 320$ minutes,
there may be a clustering of WD temperatures around 30,000K. These include
U Gem, SS Aur, WW Ceti, BD Pav, RX And, UU Aql and TU Men. The WD in the
long period SU UMa system TU Men has a temperature at the hot extreme of
the WDs in the SU UMa systems below the period gap. Above 380 minutes,
there appears to be a considerable spread in the WD T$_{eff}$ with all
having T$_{eff}$ = 40,000K or greater, typically hotter than the group
between 200 minutes and 320 minutes. This hotter group includes SS Cyg, Z
Cam, RU Peg, BV Cen and V442 Cen. Possible evidence of a classical nova
shell has been advanced for Z Cam by Shara et al. (2007), who constrain
the nova outburst to have occurred between 240 and 2400 years ago.
 
If we take our derived temperatures at face value and are mindful of the
small number of systems analyzed above the gap, then it is difficult to
convincingly argue any physical significance to the two groupings.
However, it is tempting to speculate on several possibilities. If CVs
begin their lives at long periods and evolve to shorter periods as
expected, then the larger spread of $T_{eff}$ at the longer periods is
possibly manifesting the fact that their long term core-envelope thermal
coupling in response to compressional heating by time-averaged accretion
(\citep{sio95a, tow03}) has not yet achieved equilibrium. The greater
spread with $P_{orb}$ may also be due to WDs in CVs having a wider range
of core temperatures at the onset of CV evolution.  That is, it may be
possible that some dwarf novae began their mass transfer while still quite
hot. They might have become CVs before their cores had cooled to
$10^{7}$K. On the other hand, their WDs may be less massive and hence
cooler after the same degree of long term accretion. It is also possible
that systems at very long periods with evolved secondaries have a
different evolutionary history than shorter period dwarf novae. It is also
not unexpected that the long term, time-averaged accretion rates are
variable.

There may be one important difference apparent between the
ditributions of CV WD temperatures above the period gap versus below the
period gap. Namely, the dispersion in CV WD temperatures above the period
gap appear to be substantially greater than one finds below the period gap where
there is a surprisingly narrow dispersion in temperatures around 15,000K.
In order to better illustrate this difference, we have plotted in figure 6
the temperatures of the white dwarfs in non-magnetic CVs during nova-like
low states and dwarf novae quiescences (filled triangles) and magnetic CVs
during polar low states (filled circles). All of the temperatures except
for seven non-magnetic CVs above the period gap are taken from the
compilation in Table 7 of Araujo-Betancor et al. (2005). The three objects
close to the 3 hour upper boundary of the period gap with WD temperatures
above 40,000K are all VY Sculptoris nova-like variables, TT Ari, MV Lyra
and DW UMa. While the difference between the two distributions looks real,
due consideration of the smaller number of Teff values above the period
gap compared with below the period gap as well as the errors in individual
temperatures make any such conclusion tentative. 

Clearly, further FUV spectroscopic observations of more systems and new
generations of synthetic spectral and evolutionary accretion modeling will
be needed to elucidate the evolutionary history and future of these
systems.

\acknowledgements
This work was supported by {\it{HST}} snapshot
grants GO-09357.02A, GO-09724.02A by NSF grant AST05-07514 (EMS), NASA
grant NNG04GE78G (EMS), and HST snapshot grant GO-09724.06A (PS) . DdM
acknowledges financial support by ASI and INAF via grant I/023/05/0. The
ISM model we used in this work was modeled by P.E.Barrett (USNO) in the
context of the analysis of FUSE spectra of some NLs related to a different
project (PI Godon).  PG wishes to thank Mario Livio at 
the Space Telescope Science Institute for his kind hospitality.

\clearpage

\figcaption{
The best-fitting single temperature WD fit to the {\it{HST}} STIS spectrum
of TU Men, the longest period SU UMa-type dwarf nova. The white dwarf
model has temperature of 28,000K for a distance of 288 pc. The model metal line profiles
are a bit broader than the observed profiles implying $v_{rot} \sin i = 400
$km/s. The fit to the continuum and Ly$\alpha$ profile
is quite reasonable. 
}

\figcaption{
A WD plus accretion disk fit to the {\it{HST}} STIS spectrum of BD Pav.
The best-fitting combination consisted of a 21,000K white dwarf with Log
$g= 9$ (the dotted line), and an accretion disk model (the dashed line) with M$_{wd} = 1 M_{\sun}$, $i =
75^{o}$, $\dot{M} = 3.0 \times 10^{-11} M_{\sun}$/yr for a distance of 500 pc . The white dwarf
With this combination (the solid line), the white dwarf provides 23\% of the flux and the accretion
disk contributes 77\% of the FUV flux.
}

\figcaption{
A WD best-fit to the {\it{HST}} STIS spectrum of SS Aur for d = 200 pc.
The best-fitting WD model has $34,000 \pm 2000$K with Log
$g= 8.8$.}

\figcaption{ 
a.
The best-fitting single temperature WD fit
compared to the {\it{HST}} STIS spectrum (E(B-V)=0) 
of the dwarf nova TT Crt.  The best-fit
yielded T$_{eff} = 29,000$K$\pm2000$K, log $g = 8.5$, $V \sin i = 200$
km/s$\pm 100$ km/s with a distance of 506 pc. 
\\ 
b. 
The best-fitting combination WD + optically thick model disk
(solid line) compared to the {\it{HST}} STIS spectrum (E(B-V)=0) 
of the dwarf nova TT Crt.  The best-fit
yielded for the WD (the dotted line), T$_{eff} = 30,000$K$\pm2000$K, log $g = 8.7$, 
$v_{rot} \sin i = 200$km/s$\pm 100$ km/s, and for the accretion disk (the dashed line), a 
disk inclination angle of $60^{o}$, $\dot{M} = 3.0 \times 10^{-11} M_{\sun}$/yr.
The combination model continuum shortward of 1400 \AA\ is too low.}

\figcaption{
A WD fit to the {\it{HST}} STIS spectrum of V442 Cen with E(B-V) = 0.10.
The best-fitting model by far consisted of a 47,000K white dwarf with 
Log $g= 8.3$ but for a distance of only 328 pc, well below our best 
estimate of 800 pc.}

\figcaption{
WD temperatures versus orbital period. Filled triangles are non-magnetic CVs,
filled circles are  low-state magnetic CVs (polars)  with known WD temperatures}








\end{document}